\documentclass[journal,a4paper]{IEEEtran}
\usepackage{cite}

\ifCLASSINFOpdf
   \usepackage[pdftex]{graphicx}
\else
\fi
\usepackage{calc}
\usepackage[cmex10]{amsmath}
\usepackage{bbm}
\usepackage{dsfont}
\usepackage{amsfonts,latexsym,amsmath}
\usepackage{epstopdf}
\usepackage{amssymb}
\usepackage{tikz}
\usetikzlibrary{positioning,shadows,arrows}
\usepackage{caption}
\usepackage{algorithmic}

\usepackage{array}

\usepackage{bm}
\usepackage{mdwmath}
\usepackage{mdwtab}

\usepackage{eqparbox}

\usepackage[tight,footnotesize]{subfigure}

\def\b0{{\rm 0}}

\newfont{\bb}{msbm10 scaled 1100}

\newtheorem{thm}{Identity}

\begin{document}

\newcommand{\HH}[1]{{\color{blue}#1}}

\title{Optical Fiber MIMO Channel Model and its Analysis}

\author{Apostolos~Karadimitrakis,
        Aris~L.~Moustakas, Hartmut~Hafermann and Axel~Mueller
\thanks{A. Karadimitrakis (apokaradim[at]phys.uoa.gr) 
and A. L. Moustakas (arislm[at]phys.uoa.gr) are with the Department
of Physics, University of Athens, Greece. H. Hafermann (hartmut.hafermann[at]huawei.com) and A. Mueller (axel.mueller[at]huawei.com) are with Mathematical and Algorithmic Sciences Lab, France Research Center, Huawei Technologies Co. Ltd.  }}

\maketitle

\begin{abstract}
\boldmath

Technology is moving towards space division multiplexing in optical fiber to keep up the trend in rate increase over time and to avoid an imminent capacity crunch. Thus, it is of paramount interest to estimate the potential gains of this approach.
As more spatial channels are being packed into a single fiber, the increased crosstalk necessitates the use of MIMO to guarantee reliable operation. In this paper, we exploit the analogy between an optical fiber and a model from mesoscopic physics -- a chaotic cavity -- to obtain a novel channel model for the optical fiber. The model captures both random distributed crosstalk and mode-dependent loss, which are described within the framework of scattering theory. Using tools from replica theory and random matrix theory, we derive the capacity of the fiber optical MIMO channel model.
\end{abstract}


\IEEEpeerreviewmaketitle

\section{Introduction}

\IEEEPARstart{T}{he} enormous amount of information produced by everyone in their everyday life is characteristic of our modern society. In recent years, there has been a significant change in the pattern of how information is exchanged. Previously information was largely produced by few big entities (e.g. news sites, entertainment organizations, etc.) and downloaded by individuals. More recently, 
individual users have been transformed into hubs themselves and are now a large source of information; an always-connected entity.
This changes the structure of the network and increases traffic demands.
Additionally new services, in particular video streaming, put an increasing load on today's infrastructure, thus taking it to its limits and forcing us to find a solution to expand the throughput capacity and meet the increasing demand. A large fraction of this load is carried by optical fiber networks, which form the backbone of the internet and other communication networks. As technology approaches the physical limits of single mode fiber and the capacity reaches its practical limits of the order of $100$Tbit/s, a capacity crunch is imminent~\cite{chralyvy2009plenary}.

\begin{figure}
	\centering{%
		\includegraphics[scale=0.4]{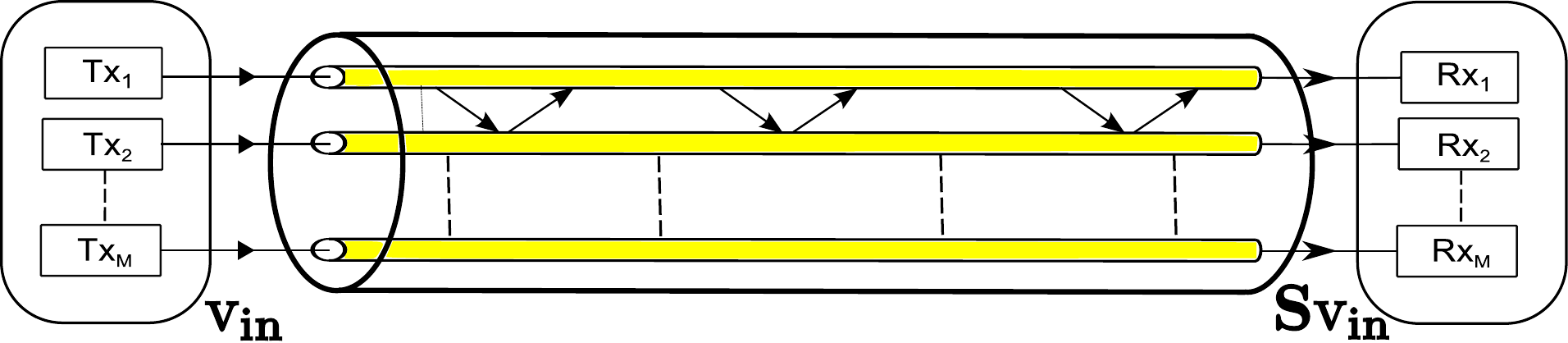}
		\caption{Illustration of crosstalk between spatial channels in optical fiber with input power \textbf{\textrm{v}}$_{in}$ and output \textbf{\textrm{v}}$_{out} = \mathbf{S}$\textbf{\textrm{v}}$_{in}$.}
		\label{xtalk}
		\vspace{-1.5em}
	}
\end{figure}

A candidate technology to avoid such a scenario and to keep up the trend of roughly ten-fold increase in capacity every four years, is space division multiplexing (SDM)~\cite{Richardson2013}. SDM utilizes space as the final remaining degree of freedom, by tightly packing several spatial channels into a single fiber. This can be done using multiple cores in multicore fibers (MCF) or multiple modes in multimode fibers (MMF).
In MMF, orthogonal modes are coupled due to fiber imperfections or twisting and bending. Crosstalk between modes cannot be avoided in long-haul transmission.
In MCF, crosstalk levels can be kept sufficiently low to be negligible only for moderate number of cores. However, it may be beneficial to deliberately introduce crosstalk, as it gives rise to supermodes with higher effective mode area. This implies a reduced impact of nonlinear impairments and enables higher channel densities.. Such fibers also exhibit a sublinear scaling of mode group dispersion with distance~\cite{Ryf2012b}. Mode scrambling is also introduced by certain types of nearly lossless multiplexers, so-called photonic lanterns~\cite{Birks2015}. Figure~\ref{xtalk} illustrates crosstalk in optical fiber.

In all these cases the original signals have to be disentangled at the receiver. Due to similarities with the wireless channel, the MIMO technique has been considered for this purpose.
However, the two channels also exhibit fundamental differences. For example, the optical channel does not suffer from information loss due to the absence of diffuse scattering. Backscattering is also negligible. Due to the low in-fiber power loss,
the channel matrix is still unitary, albeit with complex random elements, and the channel matrix is then a GUE (Gaussian Unitary Ensemble) with Dyson index $\beta =2$.

Previously~\cite{karadimitrakis2014outage}, we considered the outage capacity of the fiber optical MIMO channel in the limit of full subchannel mixing and in the absence of Mode Dependent Loss (MDL). 
In this paper, we move towards a more realistic model of the optical MIMO channel by establishing its analogy with a model from mesoscopic physics: The chaotic  cavity~\cite{Beenakker1997_RMT}. This model is amenable to a random matrix theory analysis and can interpolate from zero to strong mixing between subchannels and includes MDL. To showcase its validity we compute the channel's mutual information via a saddle point analysis.

We give the channel description in Section~II. In Section~III and Section~IV we present the mathematical analysis according to random matrix and replica theory and provide our numerical results. Finally, in Section~V we conclude.

\subsection{Notation}

\subsubsection{Matrices and Vectors.} We use upper case letters in bold font to denote matrices, $ \mbox{e.g.}~\mathbf{X}$, with entries given by $X_{ab}$. The superscript $T$ denotes the transpose operation, $\dagger$ denotes the conjugate transpose and $\mathbf{I}_N$ represents the $N$-dimensional identity matrix.

\subsubsection{Integral Measures} We deal with integrals over real matrix elements. We integrate over the elements of an $m_{rows} \times m_{cols}$ matrix $\mathbf{X}$. The corresponding integral measure is denoted by
\begin{eqnarray}
D\mathbf{X} \equiv \prod_{i=1}^{m_{rows} }\prod_{j=1}^{m_{cols}}d X_{ij}.
\end{eqnarray}

\section{Channel Description}
			
The optical fiber may be viewed as a cavity where optical power may enter and exit from both ends. The output power $\mathbf{v}_{\text{out}}$ is related to the input power $\mathrm{v}_{\text{in}}$ through $\mathbf{\mathbf{v}}_{out} = \mathbf{S}\mathbf{\mathbf{v}}_{in}$ (see Fig.~\ref{xtalk}) with the $2N\times2N$ scattering matrix $\mathbf{S}$

	\begin{eqnarray}
	\mathbf{S}=
	\begin{bmatrix}
	\mathbf{r}_{\ell \rightarrow \ell}&\mathbf{t}_{r \rightarrow \ell} \\
	\mathbf{t}_{\ell \rightarrow r}&\mathbf{r}_{r \rightarrow r}
	\end{bmatrix}.
	\end{eqnarray}
	The $\mathbf{t}$ and $\mathbf{r}$ submatrices correspond to the reflected and transmitted signals, respectively.
In contrast to a general cavity, backscattering in the fiber is negligible. Thus we have $\mathbf{r}_{\ell , r \rightarrow \ell , r}=0$. In addition, $\mathbf{t}_{r \rightarrow \ell} = \mathbf{t}_{\ell \rightarrow r}^{\dagger}$ because the two fiber ends are not distinguishable, therefore a signal entering from the left and propagating to the right experiences the same phenomena as a signal entering from the right end and propagates to the left.

The fiber exhibits random distributed crosstalk between modes or cores. We assume this mixing to be random over different frequency bands, due to strong delay spread. The situation is analogous to that of a chaotic cavity, which randomly mixes the cavity states.
The analytic expression of the $2N \times 2N$ scattering matrix for a chaotic  cavity reads~\cite{Beenakker1997_RMT}: 
\begin{eqnarray}
\mathbf{S}&=&\mathbf{I}-2\pi i\mathbf{W}^{\dagger}(\boldsymbol{\mathcal{H}}+i\pi\mathbf{W}\mathbf{W}^{\dagger})^{-1}\mathbf{W}.
\label{Eq: Beenakker}
\end{eqnarray}
Here $\boldsymbol{\mathcal{H}}$ is the $2 N \times 2N$  channel Hamiltonian and $\mathbf{W}$ is a $2N\times 2N$ matrix containing the coupling constants of the fiber to the outside world. The dimension is $2 N \times 2N$ as there are $N$ incoming states from the left and $N$ incoming states from the right, while inside the fiber there are $N$ states propagating from left to right and $N$ states propagating from right to left. So in case of perfect (lossless) leads, $\mathbf{W}\propto \mathbf{I}_{2N}$. 

The channel Hamiltonian is 
\begin{eqnarray}
\boldsymbol{\mathcal{H}} = \begin{bmatrix}
\mathbf{0}_N & \mathbf{H}_{r \rightarrow \ell} \\ 
\mathbf{H}_{\ell \rightarrow r} & \mathbf{0}_N
\end{bmatrix}.
\end{eqnarray} 
The offdiagonal sub-matrices vanish due to absence of reflection. $\mathbf{H}_{(\ell,r \rightarrow r,\ell) N}$ are $N \times N$ Hermitian. Because $\mathbf{H}_{r \rightarrow \ell} = \mathbf{H}_{\ell \rightarrow r}^{\dagger}$ we can write for simplicity $\mathbf{H} \equiv \mathbf{H}_{\ell \rightarrow r} $.
\eqref{Eq: Beenakker} then becomes 

 \begin{eqnarray}
\mathbf{S} = \mathbf{I}_{2N} - 2 \alpha\pi  i  \mathbf{W}^{\dagger}\left(\boldsymbol{\mathcal{H}} +i\alpha\pi\mathbf{W}\mathbf{W}^{\dagger}\right)^{-1}\mathbf{W}.
\end{eqnarray}
But, as $\mathbf{WW}^\dagger = \alpha \mathbf{I}_{2N} \ll 1$, for simplification it is $\boldsymbol{\mathcal{H}}+i\pi\mathbf{W}\mathbf{W}^{\dagger} \approx \boldsymbol{\mathcal{H}}$ and finally we have
\begin{eqnarray}
\mathbf{S} = \mathbf{I}_{2N} - 2 \alpha\pi  i  \mathbf{W}^{\dagger}\boldsymbol{\mathcal{H}} ^{-1}\mathbf{W}.
\label{Eq: Beenakker_2}
\end{eqnarray}
To model the MDL we add the $2N \times 2N$ loss matrix $\mathtt{\Gamma}$ \cite{fyodorov2005scattering} 
\begin{eqnarray}
\boldsymbol{\mathtt{\Gamma}} = \begin{bmatrix}
\mathbf{0}_N & \mathbf{\Gamma}_{r \rightarrow \ell} \\ 
\mathbf{\Gamma}_{\ell \rightarrow r} & \mathbf{0}_N
\end{bmatrix}.
\end{eqnarray}
Just as for $\boldsymbol{\mathcal{H}}$ we have $\mathbf{\Gamma} \equiv \mathbf{\Gamma}_{\ell \rightarrow r} $.
For simplicity we can assume that $\mathbf{\Gamma}$ is a diagonal matrix.

\begin{eqnarray} 
\mathbf{S} =\mathbf{I} - 2\alpha\pi i  \mathbf{W}^{\dagger}\left(\boldsymbol{\mathcal{H}} + i\mathtt{\Gamma}   \right)^{-1}\mathbf{W}
\label{Eq: Beenakker_2}
\end{eqnarray}
or
\begin{eqnarray}
\mathbf{S} &=&\mathbf{I}- 2\alpha \pi i \mathbf{W}^{\dagger}\left(\boldsymbol{\mathcal{H}} + i \mathtt{\Gamma}\right)\left( \boldsymbol{\mathcal{H}}^2 +\mathtt{\Gamma} ^2 \right)^{-1}\mathbf{W}.
\end{eqnarray}

\subsection{Statement of Problem}

We wish to compute the capacity of the optical MIMO channel. The mutual information is given by the well-known expression
\begin{eqnarray}\label{Eq:mutual_info}
	\mathcal{I}(\mathbf{y};\mathbf{x}|\mathbf{U}) = \left\langle\log \mbox{det }(\mathbf{I}+\rho _0\mathbf{U}\mathbf{U}^{\dagger})\right\rangle,
	\end{eqnarray}
	where $\rho _0$ is the signal strength, $\mathbf{U}$ is the complex $N_r\times N_t$ channel matrix where $N = N_t+ N_r$ and $N_t$, $N_r$ are the number transmitted and reflected modes. $\mathbf{x,y}$ are $N_t$ and $N_r$ dimensional vectors of the transmitted and received signals,  respectively. Both are assumed to be zero-mean Gaussian. The maximum of the mutual information over the input distribution yields the capacity of the channel. The capacity is the maximum error-free information transmission rate when the channel matrix $\mathbf{U}$ varies through its whole distribution $p(\mathbf{U})$. We assume $\mathbf{U}$ to be Gaussian distributed.

The $N_t\times N_r$ matrix $\mathbf{U}$ is a sub-matrix of the $2N \times 2N$ matrix $\mathbf{S}$. To extract $\mathbf{U}$ we use two diagonal $2N\times 2N$ matrices $\mathbf{A}_{\text{diag}}$ and $\mathbf{B}_{\text{diag}}$ with $\mathbf{A}_{\text{diag}}\mathbf{B}_{\text{diag}}=0$.
	\begin{eqnarray}
	\mathbf{U} = \mathbf{A}_{\text{diag}}\mathbf{S}\mathbf{B}_{\text{diag}}
	\end{eqnarray}
and
	\begin{eqnarray}
	\mathbf{U}^{\dagger} = \mathbf{B}_{\text{diag}}^\dagger\mathbf{S}^\dagger\mathbf{A}_{\text{diag}}^\dagger,
	\end{eqnarray}
	which yields
	\begin{eqnarray}
	\mathbf{U}^\dagger\mathbf{U} = \mathbf{B}_{\text{diag}}^\dagger\mathbf{S}^\dagger\mathbf{A}_{\text{diag}}^\dagger\mathbf{A}_{\text{diag}}\mathbf{S}\mathbf{B}_{\text{diag}},
	\end{eqnarray}
	so that we can write, 
	\begin{eqnarray}
	\mathbf{U}^\dagger\mathbf{U} &=&  2\alpha i  \pi\left(\mathbf{H} - i \mathbf{\Gamma}\right)\left( \mathbf{H}^2 +\mathbf{\Gamma} ^2\right)^{-1}\times\nonumber\\
	&&\left(-   2\alpha i \pi \left(\mathbf{H} + i \mathbf{\Gamma}\right)\left( \mathbf{H}^2 +\mathbf{\Gamma} ^2 \right)^{-1}\right)\nonumber\\
	&=&4\alpha^2 \pi ^2 \left(\mathbf{H}^2 +\mathbf{\Gamma} ^2 \right)^{-1}.
	\label{UU}
	\end{eqnarray}
	To incorporate the idea of a fading channel subject to crosstalk in our analysis, we assume that the channel $\mathbf{H}$ consists of a random part $\mathbf{G}$, plus a deterministic part $\mathbf{H}_0$. Thus the final equation becomes
\begin{eqnarray}
\mathbf{U}^{\dagger}\mathbf{U} &=& 4 \alpha ^2\pi ^2 \left((\mathbf{H}_0+\gamma\mathbf{G})^2 +\mathbf{\Gamma} ^2\right)^{-1}.
\label{Eq:UU}
\end{eqnarray}	
Here $\gamma$ is a parameter controlling the randomness. $\mathbf{H}_0$ is a diagonal matrix and corresponds to the line-of-sight component inside the fiber while the Gaussian distributed matrix $\mathbf{G}$ describes the crosstalk.

\section{Analysis}
In the previous sections we introduces the channel. Next, we will showcase the behavior of the mutual information and we will derive our results.
\subsection{Replica Theory}
We start from the mutual information Eq.~\eqref{Eq:mutual_info} of an optical MIMO channel.
By introducing Eq.~\eqref{Eq:UU} to Eq.~\eqref{Eq:mutual_info} we obtain
\begin{eqnarray}
	\mathcal{I}(\mathbf{y};\mathbf{x}|\mathbf{H}_0,\mathbf{G}) =\nonumber\\
	 \left\langle\log \mbox{det }\left(\mathbf{I}+\rho _04\alpha ^2 \pi^2 \left((\mathbf{H}_0+\gamma\mathbf{G})^2  + \mathbf{\Gamma}^2\right)^{-1}\right)\right\rangle\nonumber\\
	=\bigg\langle\log\mbox{det}\left[(\mathbf{H}_0 +\gamma\mathbf{G})^2 +\mathbf{\Gamma}^2 + \rho \mathbf{I}\right]- \nonumber\\ -\log\mbox{det}\left[(\mathbf{H}_0+\gamma\mathbf{G})^2+\mathbf{\Gamma}^2\right]\bigg\rangle \nonumber\\
	=\mathbb{E}\left[\mathcal{I}_1 - \mathcal{I}_2\right],
	\label{Eq:I}
	\end{eqnarray}
where 
\begin{eqnarray}
\rho &=& 4 \alpha ^2\rho _0 \pi ^2, \\ \mathbf{F} &=&  \mathbf{\Gamma}^2 + \rho\mathbf{I}, \\
\mathcal{I}_1 &=&  \log\mbox{det}\left[(\mathbf{H}_0 +\gamma\mathbf{G})^2 +\mathbf{F},\right]\\
\mathcal{I}_2&=& \log\mbox{det}\left[(\mathbf{H}_0+\gamma\mathbf{G})^2+\mathbf{\Gamma}^2\right].
\end{eqnarray}

The generating function of \eqref{Eq:mutual_info}, following \cite{Moustakas03MIMOCap}, is 
\begin{eqnarray}
g(\nu)&=&\left\langle\left[\mbox{det}\left(\mathbf{I}+\gamma _0\mathbf{U}^{\dagger}\mathbf{U}\right)\right]^{-\nu}\right\rangle = \left\langle e^{-\nu\mathcal{I}}\right\rangle\nonumber\\
&=&1-\nu \langle\mathcal{I}\rangle+\frac{\nu^2}{2}\langle \mathcal{I}^2 \rangle+\dots
\end{eqnarray}

 So we have
\begin{eqnarray}
g({\nu _1,\nu _2}) &=&\left\langle e^{-(\nu _1 \mathcal{I}_1 + \nu _2 \mathcal{I}_2)}\right\rangle,
\label{Eq:g_I}
\end{eqnarray}
We are interested in the mean $ \mathcal{I}= \left\langle \mathcal{I}_1\right\rangle - \left\langle \mathcal{I}_2\right\rangle $ and the variance $\text{var}\left(\mathcal{I}\right) = \text{var}\left(\mathcal{I}_1\right) + \text{var}\left(\mathcal{I}_2\right) - 2\text{covar}\left( \mathcal{I}_1, \mathcal{I}_2 \right) $ of the mutual information. Therefore we have to calculate the mean value of both $\mathcal{I}_1$ and $\mathcal{I}_2$ and also their respective variances and their covariance. In order to set ourselves either to $\nu _1$- or $\nu _2$-space, we set $\nu _2 = 0$ or $\nu _1 =0$, respectively. The $N \rightarrow \infty $ and $ \nu \rightarrow 0^+$ limits in the evaluation of $g(\nu) $ can be interchanged by first taking the former and then the latter without changing the final answer. Indeed, the two limits of large number of propagating modes and small $\nu$ are not different from each other. Higher terms in the $\nu$  expansion lead to higher terms in the $\frac{1}{N}$ expansion.

\subsubsection{Calculation of  $ {g_{\mathcal{I}_1}(\nu _1)}  $}
\hfill\break

Using Identity \ref{iden:1} (Appendix~\ref{ids}), we can write \newline $g_{\mathcal{I}_1}(\nu _1) = \left\langle \mbox{det} \left[\left(\mathbf{H}_0 + \gamma\mathbf{G}\right)^2 + \mathbf{F}\right]^{-\nu _1}\right\rangle $, as

\begin{eqnarray}
g_{\mathcal{I}_1}(\nu _1) &=& \int D\mathbf{X}e^{-\frac{1}{2}\text{Tr}\{\mathbf{X}^{\dagger}\left(\left(\mathbf{H}_0 + \gamma\mathbf{G}\right)^2 + \mathbf{F}\right)\mathbf{X}\}}\\
&=& \int D\mathbf{X}e^{-\frac{1}{2}\text{Tr}\{\mathbf{X}^{\dagger}\mathbf{F}\mathbf{X}\}}\langle e^{-\frac{1}{2}\text{Tr}\{\mathbf{X}^{\dagger}\left(\mathbf{H}_0 + \gamma\mathbf{G}\right)^2\mathbf{X}\}} \rangle _\mathbf{G}\nonumber.
\end{eqnarray}
Using Identity~\ref{iden:2} setting $\mathbf{A}^{\dagger} = -i\mathbf{X}^{\dagger}(\mathbf{H}_0 + \gamma \mathbf{G})$ and $\mathbf{B} = i\mathbf{X}(\mathbf{H}_0 +\gamma\mathbf{G})$, we write
\begin{eqnarray}
\langle e^{-\frac{1}{2}\text{Tr}\{\mathbf{X}^{\dagger}\left(\mathbf{H}_0 + \gamma\mathbf{G}\right)^2\mathbf{X}\}} \rangle = \nonumber\\
 \int D\mathbf{Y}e^{-\frac{1}{2}\left(\mathbf{Y}^T\mathbf{Y}+ i \mathbf{Y}^T(\mathbf{H}_0 + \gamma\mathbf{G})\mathbf{X}\right) + \left(\mathbf{Y}^T\mathbf{Y}+ i \mathbf{X}^T(\mathbf{H}_0 + \gamma\mathbf{G})\mathbf{Y}\right) }.
\end{eqnarray}
All-together we obtain
\begin{eqnarray}
g_{\mathcal{I}_1}(\nu _1) = \int D\mathbf{X}\int D\mathbf{Y} e^{-\frac{1}{2}\left(\mathbf{X}^T\mathbf{F}\mathbf{X} + \mathbf{Y}^T\mathbf{Y} \right)}\times\nonumber\\
e^{-\frac{i}{2}\left(\mathbf{Y}^T\mathbf{H}_0 \mathbf{X} + \mathbf{X}^T\mathbf{H}_0 \mathbf{Y} \right)}
\langle e^{-\frac{i\gamma}{2}\left( \mathbf{X}^T\mathbf{G} \mathbf{Y} + \mathbf{Y}^T\mathbf{G} \mathbf{X}\right)}\rangle _\mathbf{G}.
\end{eqnarray}
The ensemble average over channel realization for an arbitrary function is
\begin{eqnarray}
\langle e^{-\frac{i \gamma}{2}\left( \mathbf{X}^T\mathbf{G} \mathbf{Y} + \mathbf{Y}^T\mathbf{G} \mathbf{X}\right)}\rangle _\mathbf{G} = \nonumber\\
\int D\mathbf{G} e^{-\frac{N}{2}\text{Tr}\{\mathbf{G}^2\}}e^{-\frac{i \gamma}{2}\text{Tr}\{\mathbf{G}\left(\mathbf{XY}^T + \mathbf{YX}^T\right)\}}\propto\nonumber\\
e^{-\frac{1}{2}\frac{\gamma ^2}{4 N}\left( \mathbf{XY}^T + \mathbf{YX}^T \right)^2}.
\end{eqnarray}
The last exponential can be written as 
\begin{eqnarray}
e^{\frac{\gamma}{8N}\mbox{Tr}\left(2(\mathbf{X}^T\mathbf{X}\mathbf{Y}^T\mathbf{Y}) + (\mathbf{X}^T\mathbf{Y} )^2 +(\mathbf{Y}^T\mathbf{X})^2\right)}.
\label{Eq:HST1}
\end{eqnarray}
In order to evaluate the first term of \eqref{Eq:HST1} we will use Identity \ref{iden:3} and introduce $\nu _1\times \nu _1$ matrices $\mathcal{R, T }$: 
\begin{eqnarray}
e^{\frac{\gamma}{8N}\mbox{Tr}\left(2(\mathbf{X}^T\mathbf{X}\mathbf{Y}^T\mathbf{Y})\right)} &=& \nonumber\\\int D(\mathcal{T,R}) e^{N\mbox{Tr}\left(\mathcal{TR}\right)}e^{-\frac{\gamma}{2}\mbox{Tr}\left( \mathbf{Y}^T\mathbf{Y}\mathcal{R}+\mathcal{T}\mathbf{X}^T\mathbf{X}\right)}.
\end{eqnarray}
The evaluation of the quadratic parts of the exponential \eqref{Eq:HST1} is more tricky. This time we will introduce  $\nu _1\times \nu _1$ matrices $\mathcal{P, Q }$. We have
\begin{eqnarray}
e^{\frac{\gamma}{8N}\mbox{Tr}\left((\mathbf{X}^T\mathbf{Y} )^2 +(\mathbf{Y}^T\mathbf{X})^2\right)} =\nonumber\\
\int D\mathcal{P} e^{-N\mathcal{P}^2}e^{-\frac{i\gamma}{2}\mbox{Tr}\left(\mathcal{P}(\mathbf{X}^T\mathbf{Y}+ \mathbf{Y}^T\mathbf{X})\right)} \nonumber\\
 +\int D\mathcal{Q} e^{-N\mathcal{Q}^2}e^{-\frac{i\gamma}{2}\mbox{Tr}\left(\mathcal{Q}(\mathbf{X}^T\mathbf{Y}- \mathbf{Y}^T\mathbf{X})\right)}.
\end{eqnarray}

\paragraph{Saddle-point analysis}
So, bringing everything together, we have
\begin{eqnarray}
g_{\mathcal{I}_1}(\nu _1) = \int D(\mathcal{T, R, P, Q})e^{-\mathcal{S}},
\end{eqnarray}
where
\begin{eqnarray}
\mathcal{S} &=& -N\mbox{Tr}\left( \mathcal{TR} -\mathcal{P}^2 - \mathcal{Q}^2\right)\nonumber\\
&& +\log\mbox{det}\begin{bmatrix}
\mathbf{F} + \gamma\mathcal{T} & i \left(\mathbf{H}_0 +\gamma (\mathcal{P +Q})  \right) \\ 
i \left(\mathbf{H}_0 +\gamma (\mathcal{P -Q})  \right) & 1 + \gamma \mathcal{R}
\end{bmatrix}.\nonumber\\
\label{Eq:S} 
\end{eqnarray}
To consider the vicinity near the saddle-point we rewrite $\mathcal{R,T,P,Q}$ as

\begin{eqnarray}
\mathcal{T} &=&  t\mathbf{I}_{\nu _1} + \delta \mathbf{T} \nonumber\\
\mathcal{R} &=&r\mathbf{I}_{\nu _1} + \delta \mathbf{R} \nonumber\\
\mathcal{P} &=&  p\mathbf{I}_{\nu _1} + \delta \mathbf{P} \nonumber\\
\mathcal{Q} &=&  q\mathbf{I}_{\nu _1} + \delta \mathbf{Q},
\label{Eq:TRPQ}
\end{eqnarray}

where $\delta\mathbf{T},\delta\mathbf{R},\delta\mathbf{P},\delta\mathbf{Q}$ are $\nu _1 \times \nu _1 $ matrices which represent deviations around the saddle point.

That way, we can use the Taylor expansion for $\mathcal{S}$ of \eqref{Eq:S} as :
\begin{eqnarray}
\mathcal{S} = \mathcal{S}_0 + \mathcal{S}_1 + \mathcal{S}_2 + \mathcal{S}_3 + \dots 
\end{eqnarray}
with
\begin{eqnarray}
\mathcal{S}_0 &=&- N\mbox{Tr}(tr - p^2 - q^2) \nonumber\\
&&+ \log\bigg[ \left(\mathbf{F} + \gamma t \right)\left(\mathbf{I}+\gamma r \right)\nonumber\\
&&+\left(\mathbf{H}_0+\gamma(p-q)\right)\left(\mathbf{H}_0+\gamma(p+q)\right)  \bigg].
\end{eqnarray}
 
Continuing the evaluation of the next term of the Taylor expansion, since we are looking for saddle point solution, $\mathcal{S}$ must be stationary with respect to variations in $\mathcal{R,T,P,Q}$. Therefore $\mathcal{S}_1 =0$ and the corresponding saddle-point equations are:

\begin{eqnarray}
r &=& \frac{1}{N}\text{Tr} \frac{\gamma(1 + \gamma r)}{\left(\mathbf{F} + \gamma t \right)\left(\mathbf{I}+\gamma r \right) + (\mathbf{H}_0  - \gamma p)^2 }\\
p &=&  \frac{1}{N}\text{Tr} \frac{\gamma(\mathbf{H}_0 - \gamma p)}{\left(\mathbf{F} + \gamma t \right)\left(\mathbf{I}+\gamma r \right) + (\mathbf{H}_0  - \gamma p)^2 }\\
t &=& \frac{1}{N}\text{Tr} \frac{\gamma(\mathbf{F} + \gamma t)}{\left(\mathbf{F} + \gamma t \right)\left(\mathbf{I}+\gamma r \right) + (\mathbf{H}_0  - \gamma p)^2 }\\ 
q &=& 0
\end{eqnarray}
and the second order term is  
\begin{eqnarray}
\mathcal{S}_2 &=& \frac{1}{2} \text{Tr} \left\lbrace \begin{bmatrix}
\delta\mathbf{T} \\ 
\delta\mathbf{R} \\ 
\delta\mathbf{P} \\ 
\delta\mathbf{W}
\end{bmatrix} ^T \mathbf{\Sigma} \begin{bmatrix}
\delta\mathbf{T} \\ 
\delta\mathbf{R} \\ 
\delta\mathbf{P} \\ 
\delta\mathbf{W}
\end{bmatrix} \right\rbrace ,
\label{Eq:Variance}
\end{eqnarray}
where $\mathbf{\Sigma}$ is a $4\times 4$ Hessian matrix the entries of which can be seen in Appendix~\ref{HessianMatrix}.

Following \cite{Moustakas03MIMOCap}, the final outcome for the variance is:
\begin{eqnarray}
\left\langle \mathcal{I}^2\right\rangle - \left\langle \mathcal{I}\right\rangle^2 = - \log \text{det} | \mathbf{\Sigma}| \,.
\end{eqnarray}

\subsubsection{Calculation of $g_{\mathcal{I}_2}(\nu)$}
For the calculation of the $\left\langle\mathcal{I}_2\right\rangle$ and the corresponding variance, we follow the same method as above but instead of $\mathbf{F}$ we only have $\mathbf{\Gamma}^2$. 

\subsubsection{Calculation of $g_{\mathcal{I}_{12}}(\nu _1, \nu_2)$ (Covariance)}

Again, using Identity 1

\begin{eqnarray}
g_{\mathcal{I}_1\mathcal{I}_2}(\nu _1, \nu _2) &=& \int D\mathbf{X}_1D\mathbf{X}_2\nonumber\\
&&\Bigg\langle \text{exp}\Bigg({-\frac{1}{2}\mathbf{X}_1^\dagger\left[ \mathbf{F} +\left( \mathbf{H}_0 + \gamma\mathbf{G} \right)^2\right]\mathbf{X}_1} \nonumber\\
&-&\frac{1}{2}\mathbf{X}_2^\dagger\left[ \Gamma ^2 +\left( \mathbf{H}_0 + \gamma\mathbf{G} \right)^2 \right]\mathbf{X}_2\Bigg)\Bigg\rangle.
\end{eqnarray}

Following the previous method we have
\begin{eqnarray}
g_{\mathcal{I}_1\mathcal{I}_2}(\nu _1, \nu _2) =\nonumber\\
 \int D\mathbf{X}_1D\mathbf{X}_2 \exp\Bigg({-\frac{1}{2}\text{Tr}\Bigg\{\mathbf{X}_1^\dagger\mathbf{F}\mathbf{X}_1} +\mathbf{X}_2^\dagger\mathbf{\Gamma} ^2\mathbf{X}_2\Bigg)\Bigg\}\times\nonumber\\
 \int D\mathbf{Y}_1D\mathbf{Y}_2 \exp\Bigg({-\frac{1}{2}\text{Tr}\Bigg\{\mathbf{Y}_1^\dagger\mathbf{Y}_1} +\mathbf{Y}_2^\dagger\mathbf{Y}_2 \nonumber\\
 + i\mathbf{X}_1^\dagger\mathbf{H}_0\mathbf{Y}_1 +i\mathbf{X}_2^\dagger\mathbf{H}_0\mathbf{Y}_2\Bigg)\Bigg\}\times\nonumber\\
 \left\langle \exp \left( -\frac{i \gamma}{2}\left( \mathbf{X}_1^\dagger\mathbf{G}\mathbf{Y}_1 + \mathbf{X}_2^\dagger\mathbf{G}\mathbf{Y}_2 \right) \right) \right\rangle _\mathbf{G}.
\end{eqnarray}

Again, using Identity \ref{iden:2} and Identity \ref{iden:3} we introduce the ever-helpful $\nu _1 \times \nu _1$ matrices $\mathcal{R}_1, \mathcal{T}_1 , \mathcal{P}_1, \mathcal{Q}_1$, $\nu _2 \times \nu _2$ matrices $\mathcal{R}_2, \mathcal{T}_2 , \mathcal{P}_2, \mathcal{Q}_2$  and $\nu _1 \times \nu _2$ matrices $\mathcal{R}_{12}, \mathcal{T}_{12} , \mathcal{P}_{12}, \mathcal{Q}_{12}$ and following the diagonalizing method we end up to matrix $\mathbf{A}$ which can be seen in Appendix~\ref{AMatrix}.

 At the saddle point the cross-terms $\mathcal{R}_{12},\mathcal{T}_{12}$ etc are equal to zero. So $\mathbf{A}$ becomes a block-diagonal matrix and the evaluation of the determinant ( Identity~\ref{iden:1}) is just the multiplication of these 2 blocks.   

The rest of the calculations for the computation of the covariance is straightforward and the Hessian matrix for the covariance can be seen in Appendix~\ref{Hessian_cov_Matrix}.

So finally the variance is 

\begin{eqnarray}
Var &=& - \log \text{det} | \mathbf{\Sigma }_{\mathcal{I}_1}| - \log \text{det} | \mathbf{\Sigma }_{\mathcal{I}_2}| \nonumber\\
&&+ 2\log \text{det} | \mathbf{\Sigma }_{cov}|  + 4\log 2 \,.
\end{eqnarray}

\begin{figure}
\centering{%
\includegraphics[scale=0.2]{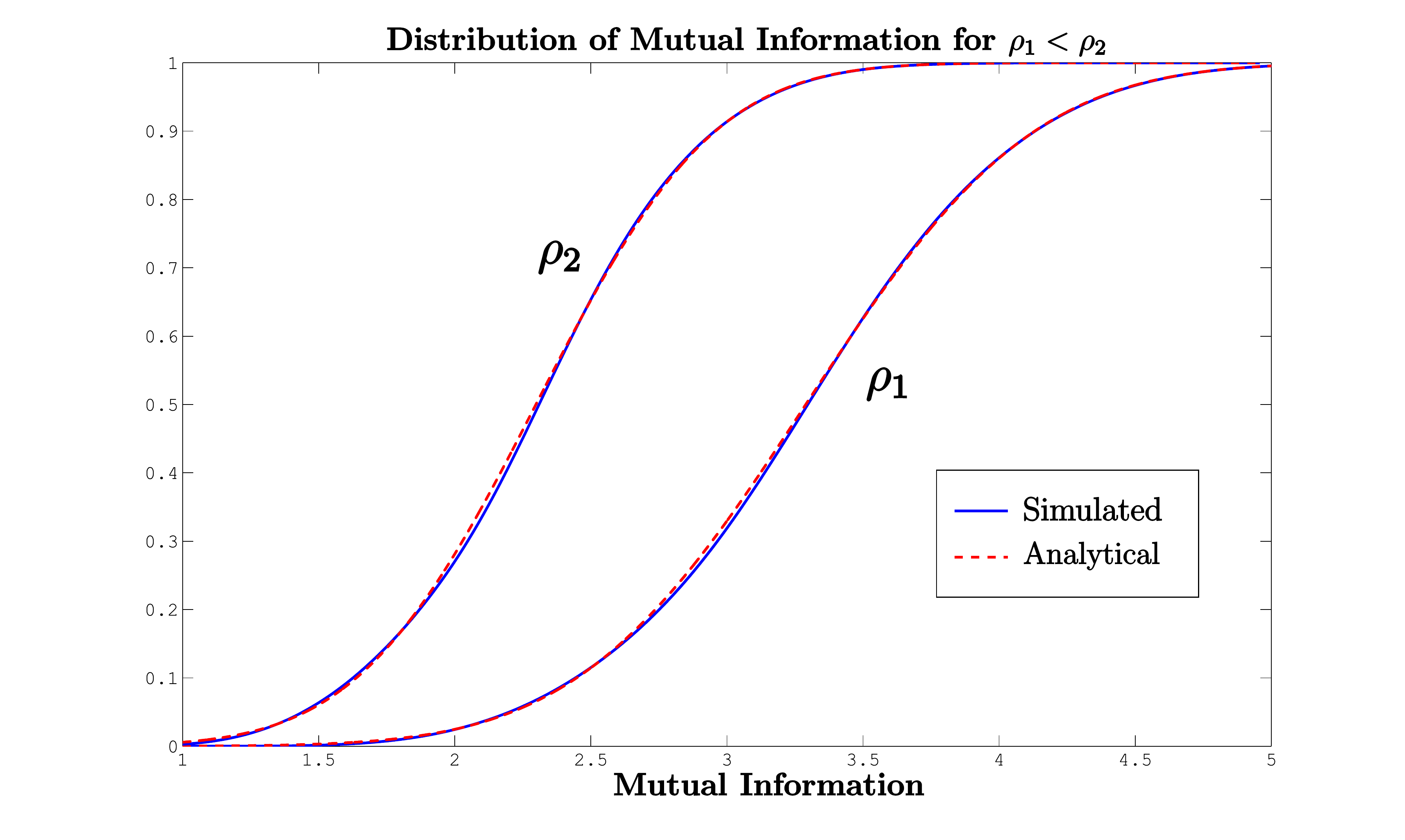}
\caption{Cumulative distribution function (CDF) of mutual information for N = 6 and $\rho _1 < \rho _2$. Runs = $10^6$}
\label{Fig:CDF}
\vspace{-1.5em}
}
\end{figure}

\begin{figure}
\centering{%
\includegraphics[scale=0.2]{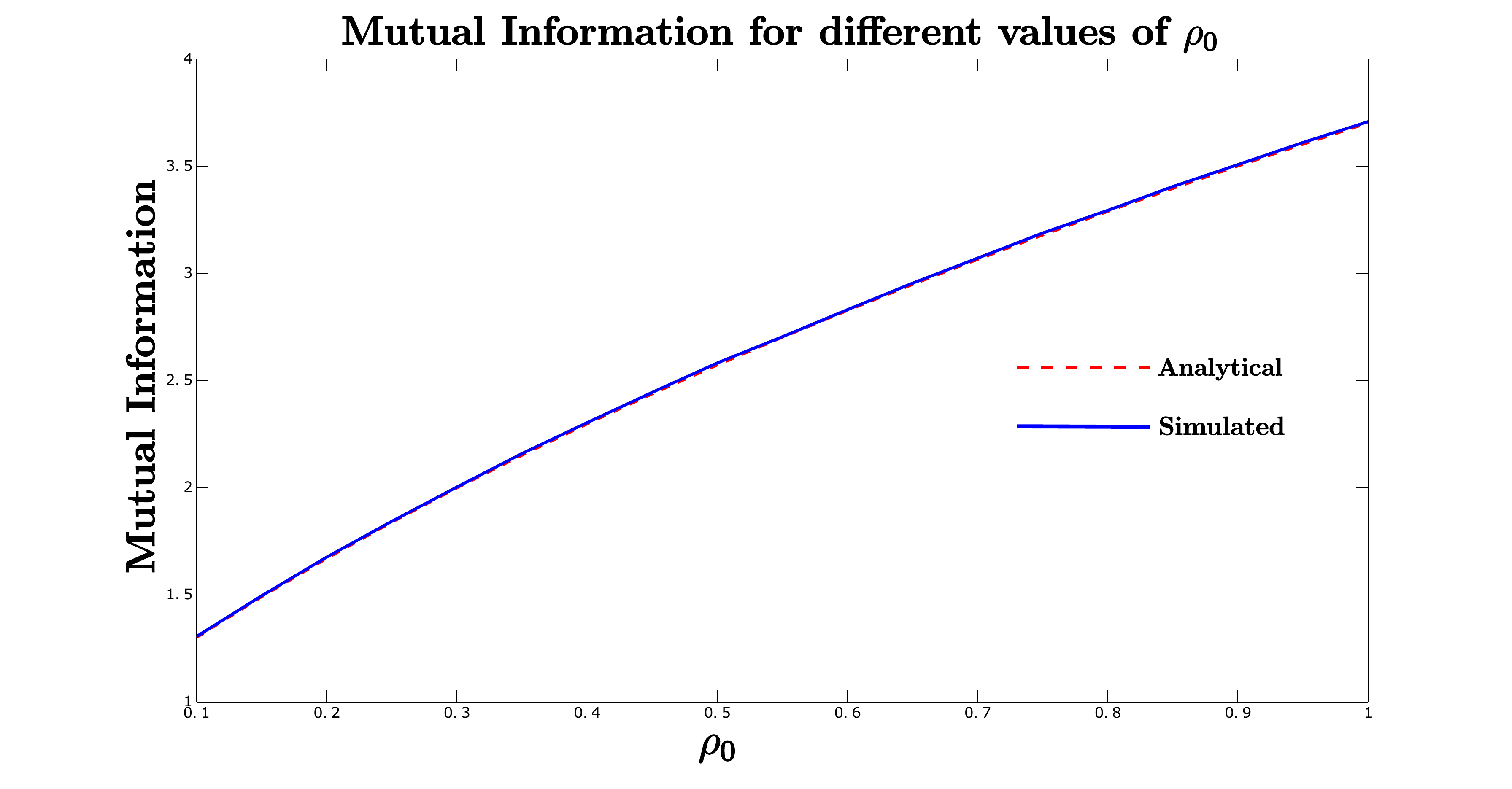}
\caption{N = 6. Runs = $10^6$}
\label{Fig:capacity}
\vspace{-1.5em}
}
\end{figure}

\section{Numerical results}
In order to check our analytical results we have numerically computed mutual information and its cumulative distribution function (CDF).
The results are shown in Figs.~\ref{Fig:CDF} and ~\ref{Fig:capacity}.
We compare the Gaussian distribution $\mathcal{N}(\langle\mathcal{I}\rangle, \text{var}(\mathcal{I}) )$ evaluated using the analytical calculations in this paper with results obtained by averaging over a large number of random matrix realizations for $N=6$.
Remarkably, we find that in both cases, the discrepancy between the different curves is small. This shows that our analytical results are valid even for the practical relevant case of small number of modes or cores $N$.

\section{Conclusions}

In this paper we have introduced a novel channel model for the optical MIMO channel by relating it to a chaotic  cavity. This new modeling approach captures the fundamental properties of the optical MIMO channel. Using tools from random matrix theory and a saddle point analysis, we have shown that in the limit of large $N$, the distribution of the mutual information approaches a Gaussian. Numerically we find that this method is valid even for low $N$. This analytic method gives us the means to analyze the statistics of throughput in the fiber optical MIMO channel in the presence of arbitrary level of crosstalk and MDL. The introduced channel model is also amenable to extensions, such as dispersion and nonlinear effects in deterministic and random part respectively.

\appendices
\section{Identities}
The proofs for the next identities , can be found in \cite{Moustakas03MIMOCap}.
\label{ids}
\begin{thm}\label{iden:1}
Let $\mathbf{M}$ be a hermitian, positive, definite square matrix $m \times m$ and $\mathbf{X}$ a complex $m\times n $ matrix, then 
\begin{eqnarray}
(detM)^{-1}=\int D\mathbf{X}e^{-\frac{1}{2}Tr\{\mathbf{X}^{\dagger}\mathbf{MX}\}}.
\end{eqnarray}
\end{thm}
\begin{thm}\label{iden:2}
Let, $\mathbf{X},\mathbf{A},\mathbf{B}$ be $m\times n $ complex matrices, then
\begin{eqnarray}
\int D\mathbf{X}e^{-\frac{1}{2}Tr\{\mathbf{X}^{\dagger}\mathbf{X}+\mathbf{A}^{\dagger}\mathbf{X}-\mathbf{X}^{\dagger}\mathbf{B}\}}=e^{-\frac{1}{2}Tr\{\mathbf{A}^{\dagger}\mathbf{B}\}}.
\end{eqnarray}
\end{thm}
\begin{thm}[Hubbard-Stratonovich transformation]\label{iden:3}
Let, $\mathbf{U,V}$ be arbitrary complex $\nu \times \nu$ matrices, where $\nu$ here is assumed to be an arbitrary positive integer. Then,
\begin{eqnarray}
e^{-Tr\{\mathbf{UV}\}} = \int D\mathbf{T}D\mathbf{R} e^{Tr\{\mathbf{RT}-\mathbf{UT}-\mathbf{RV}\}}.
\end{eqnarray}
\end{thm}

\section{The Hessian Matrix}
\label{HessianMatrix}
\begingroup 
\renewcommand*{\arraystretch}{0.5}
\setlength{\arraycolsep}{3pt}
\begin{eqnarray*}
\resizebox{1\hsize}{!}{ 
$\mathbf{\Sigma} = \begin{bmatrix}
-\mbox{Tr}\bigg(\frac{\gamma^2}{Z_1^2}(\mathbf{I} +\gamma{r})^2\bigg) & \begin{array}{c}-\mbox{Tr}\bigg(\frac{\gamma^2}{Z_1^2}(\mathbf{I} +\gamma{r})(\Delta +\gamma{t})\\ +\frac{\gamma ^2}{Z} -1\bigg)\end{array} & -\mbox{Tr}\bigg(\frac{2\gamma^2}{Z _1^2}(\mathbf{I} +\gamma{r})(\mathbf{H}_0-\gamma{p})\bigg) & 0 \\ 
\begin{array}{c}
\mbox{Tr}\bigg(\frac{\gamma^2}{Z _1^2}(\mathbf{I} +\gamma{r})(\Delta +\gamma t)\\ +\frac{\gamma ^2}{Z_1} -1\bigg) \end{array}& -\mbox{Tr}\bigg(\frac{\gamma^2}{Z_1^2}(\Delta + \gamma{t})^2\bigg) & -\mbox{Tr}\bigg(\frac{2\gamma ^2}{Z _1^2}(\Delta + \gamma{t})(\mathbf{H}_0-\gamma{p})\bigg) & 0 \\ 
 -\mbox{Tr}\bigg(\frac{2\gamma^2}{Z _1^2}(\mathbf{I} +\gamma{R})(\mathbf{H}_0-\gamma{p})\bigg) & -\mbox{Tr}\bigg(\frac{2\gamma ^2}{Z _1^2}(\Delta + \gamma{t})(\mathbf{H}_0-\gamma{p})\bigg) & \begin{array}{c}
 -\mbox{Tr}\bigg(\frac{4\gamma ^2}{Z _1^2}(\mathbf{H}_0-\gamma{p})^2\\ +\frac{2\gamma ^2}{Z _1} - 2\bigg)
\end{array}  & 0 \\ 
0 & 0 & 0 & 2N - \mbox{Tr}\bigg(\frac{2\gamma ^2}{Z_1}\bigg)
\end{bmatrix}$
}
\end{eqnarray*} 
\endgroup
where $Z _1 = \left( \mathbf{F} +\gamma{t} \right)\left( \mathbf{I} +\gamma{r}\right) + (\mathbf{H}_0 -\gamma{p})^2$. 
\section{The Matrix A}
\label{AMatrix}
\begin{eqnarray*}
\resizebox{1\hsize}{!}{$\mathbf{A}=
\begin{bmatrix}
\mathbf{F} + \gamma\mathcal{T}_1 & i\mathbf{H}_0 + i\gamma (\mathcal{P}_1+\mathcal{Q}_1) & 0 & 0 \\ 
i\mathbf{H}_0 + i\gamma (\mathcal{P}_1-\mathcal{Q}_1) & \mathbf{I}+\gamma\mathcal{R}_1 & 0 & 0 \\ 
0 & 0 & \mathbf{\Gamma}^2+\gamma\mathcal{T}_2 & i\mathbf{H}_0 + i\gamma (\mathcal{P}_2+\mathcal{Q}_2) \\ 
0 & 0 & i\mathbf{H}_0 + i\gamma (\mathcal{P}_2-\mathcal{Q}_2) &\mathbf{I}+\gamma\mathcal{R}_2
\end{bmatrix}$} 
\end{eqnarray*} 

\section{The Hessian Matrix for the Covariance}
\label{Hessian_cov_Matrix}
\begin{figure}[h!]
\centering{%
\includegraphics[scale=0.65]{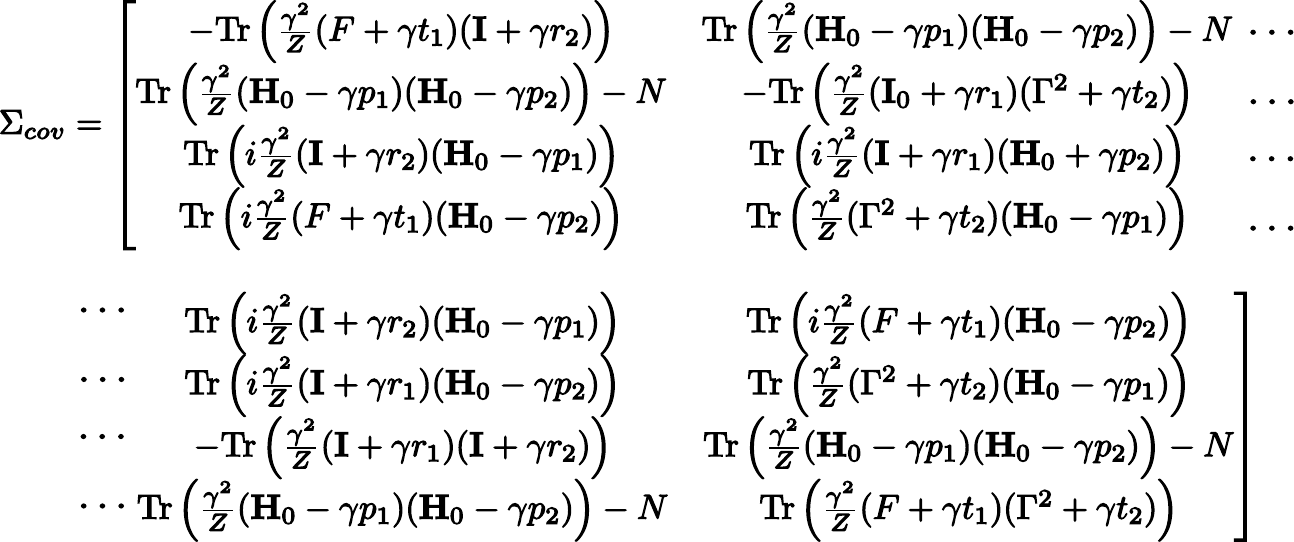}}
\end{figure}
where 
\begin{eqnarray}
Z&=& \left[ (F + \gamma t_1)(\mathbf{I}+\gamma r_1) +(\mathbf{H}_0-\gamma p_1)^2\right]\times \nonumber\\
&&\left[ (\Gamma^2 + \gamma t_2)(\mathbf{I} + \gamma r_2)+(\mathbf{H}_0 - \gamma p_2)^2\right]
\end{eqnarray}
and $F,\Gamma$ are scalars. Due to simplification reasons we assumed that the loss is not frequency selective.

\ifCLASSOPTIONcaptionsoff
  \newpage
\fi



\bibliographystyle{IEEEtran}

\end{document}